\documentclass[twocolumn, secnumarabic, amssymb, nobibnotes, aps, prb, superscriptaddress,floatfix,bibnotes,longbibliography]{revtex4-2}

\usepackage{hyperref}
\usepackage{graphicx}
 \usepackage{booktabs}
 \usepackage{subfigure} 
 \usepackage{threeparttable}
 \usepackage{xcolor}
 \usepackage{ulem}
 \usepackage{multirow}
 \usepackage{longtable}
 \usepackage{makecell}

\begin{document}

\title{
Charge Pickup Reaction Cross Section for Neutron-Rich \textit{p}-Shell Isotopes at 900\textit{A} MeV 
}

\author{J.-C. \surname{Zhang}}
\affiliation{School of Physics, Beihang University, Beijing 100191, China}
\affiliation{Research Center for Nuclear Physics (RCNP), Osaka University, Ibaraki Osaka 567-0047, Japan}

\author{B.-H. \surname{Sun}}
\email{bhsun@buaa.edu.cn}
\affiliation{School of Physics, Beihang University, Beijing 100191, China}

\author{I. Tanihata}
\email{tanihata@rcnp.osaka-u.ac.jp}
\affiliation{School of Physics, Beihang University, Beijing 100191, China}
\affiliation{Research Center for Nuclear Physics (RCNP), Osaka University, Ibaraki Osaka 567-0047, Japan}

\author{S. Terashima}
\affiliation{School of Physics, Beihang University, Beijing 100191, China}

\author{F. \surname{Wang}}
\affiliation{School of Physics, Beihang University, Beijing 100191, China}

\author{R. Kanungo}
\affiliation{Astronomy and Physics Department, Saint Mary’s University, Halifax, Nova Scotia B3H 3C3, Canada}
\affiliation{TRIUMF, Vancouver, British Columbia V6T 4A3, Canada}

\author{C. Scheidenberger}
\affiliation{GSI Helmholtzzentrum f\"ur Schwerionenforschung, D-64291 Darmstadt, Germany}
\affiliation{Justus-Liebig University, 35392 Giessen, Germany}
\affiliation{Helmholtz Research Academy Hesse for FAIR (HFHF), GSI Helmholtz Center for Heavy Ion Research, Campus Giessen, 35392 Giessen, Germany}

\author{F. Ameil}
\affiliation{GSI Helmholtzzentrum f\"ur Schwerionenforschung, D-64291 Darmstadt, Germany}

\author{J. Atkinson}
\affiliation{Astronomy and Physics Department, Saint Mary’s University, Halifax, Nova Scotia B3H 3C3, Canada}

\author{Y. Ayyad}
\affiliation{Universidad de Santiago de Compostela, E-15706 Santiago de Compostella, Spain}

\author{S. Bagchi}
\affiliation{Department of Physics, Indian Institute of Technology (Indian School of Mines) Dhanbad, Jharkhand- 826004, India}

\author{D. Cortina-Gil}
\affiliation{Universidad de Santiago de Compostela, E-15706 Santiago de Compostella, Spain}

\author{I. Dillmann}
\affiliation{GSI Helmholtzzentrum f\"ur Schwerionenforschung, D-64291 Darmstadt, Germany}
\affiliation{Justus-Liebig University, 35392 Giessen, Germany}

\author{A. Estrad\'e}
\affiliation{Astronomy and Physics Department, Saint Mary’s University, Halifax, Nova Scotia B3H 3C3, Canada}
\affiliation{GSI Helmholtzzentrum f\"ur Schwerionenforschung, D-64291 Darmstadt, Germany}

\author{A. Evdokimov}
\affiliation{GSI Helmholtzzentrum f\"ur Schwerionenforschung, D-64291 Darmstadt, Germany}

\author{F. Farinon}
\affiliation{GSI Helmholtzzentrum f\"ur Schwerionenforschung, D-64291 Darmstadt, Germany}

\author{H. Geissel}
\thanks{Deceased}
\affiliation{GSI Helmholtzzentrum f\"ur Schwerionenforschung, D-64291 Darmstadt, Germany}
\affiliation{Justus-Liebig University, 35392 Giessen, Germany} 

\author{G. Guastalla}
\affiliation{GSI Helmholtzzentrum f\"ur Schwerionenforschung, D-64291 Darmstadt, Germany}

\author{R. Janik} 
\thanks{Deceased}
\affiliation{Faculty of Mathematics and Physics, Comenius University, 84215 Bratislava, Slovakia}

\author{S. Kaur} 
\affiliation{Astronomy and Physics Department, Saint Mary’s University, Halifax, Nova Scotia B3H 3C3, Canada}
\affiliation{Department of Physics and Atmospheric Science, Dalhousie University, Halifax, Nova Scotia B3H 4R2, Canada}

\author{R. Kn\"obel}
\affiliation{GSI Helmholtzzentrum f\"ur Schwerionenforschung, D-64291 Darmstadt, Germany}

\author{J. Kurcewicz}
\affiliation{GSI Helmholtzzentrum f\"ur Schwerionenforschung, D-64291 Darmstadt, Germany}

\author{Yu. A. Litvinov}
\affiliation{GSI Helmholtzzentrum f\"ur Schwerionenforschung, D-64291 Darmstadt, Germany}

\author{M. Marta}
\affiliation{GSI Helmholtzzentrum f\"ur Schwerionenforschung, D-64291 Darmstadt, Germany}

\author{M. Mostazo} 
\affiliation{Universidad de Santiago de Compostela, E-15706 Santiago de Compostella, Spain}

\author{I. Mukha} 
\affiliation{GSI Helmholtzzentrum f\"ur Schwerionenforschung, D-64291 Darmstadt, Germany}

\author{C. Nociforo}
\affiliation{GSI Helmholtzzentrum f\"ur Schwerionenforschung, D-64291 Darmstadt, Germany}

\author{H. J. Ong}
\affiliation{Research Center for Nuclear Physics (RCNP), Osaka University, Ibaraki Osaka 567-0047, Japan}

\author{S. Pietri}
\affiliation{GSI Helmholtzzentrum f\"ur Schwerionenforschung, D-64291 Darmstadt, Germany}

\author{A. Prochazka}
\affiliation{GSI Helmholtzzentrum f\"ur Schwerionenforschung, D-64291 Darmstadt, Germany}

\author{B. Sitar} 
\affiliation{Faculty of Mathematics and Physics, Comenius University, 84215 Bratislava, Slovakia}

\author{P. Strmen} 
\thanks{Deceased}
\affiliation{Faculty of Mathematics and Physics, Comenius University, 84215 Bratislava, Slovakia}

\author{M. Takechi}
\affiliation{GSI Helmholtzzentrum f\"ur Schwerionenforschung, D-64291 Darmstadt, Germany}

\author{J. Tanaka}
\affiliation{Research Center for Nuclear Physics (RCNP), Osaka University, Ibaraki Osaka 567-0047, Japan}

\author{J. Vargas} 
\affiliation{Universidad de Santiago de Compostela, E-15706 Santiago de Compostella, Spain}

\author{H. Weick}
\affiliation{GSI Helmholtzzentrum f\"ur Schwerionenforschung, D-64291 Darmstadt, Germany}

\author{J. S. Winfield}
\thanks{Deceased}
\affiliation{GSI Helmholtzzentrum f\"ur Schwerionenforschung, D-64291 Darmstadt, Germany}

\date{\today}%

\begin{abstract} 
We report charge pickup reaction cross sections for 24 \textit{p}-shell isotopes, including $^{8,9}$Li, $^{10\text{\textendash}12}$Be, $^{10,13\text{\textendash}15}$B, $^{12,14\text{\textendash}19}$C and $^{14,15,17\text{\textendash}22}$N, measured at relativistic energies (approximately 900$A$ MeV) on both hydrogen and carbon targets.
For the first time, we reveal a universal rapid increase in the charge pickup cross sections of unstable projectiles with isospin asymmetry along several isotopic chains.
The cross sections can be decoupled into distinct contributions from the mass number and isospin asymmetry of the projectile, highlighting the critical role of the latter, and can be formulated empirically.

\end{abstract}
\maketitle

\section{Introduction}

The fragmentation of high-energy heavy ions is an important process in nuclear reactions.
Notably, it is considered the most comprehensive method for producing nuclei far from the $\beta$ stability line, enabling the creation of beams of unstable nuclei. 
The availability of such productions and beams has revived nuclear physics research and extended to astrophysical studies relevant to neutron stars and element synthesis in the Universe~\cite{bertulani2010nuclear}.

To support all those scientific works, studies of projectile fragmentation have been one of the most important subjects. 
Reaction cross sections, interaction cross sections, and charge-changing cross sections have been studied extensively to probe nuclear radii and the equation of state of asymmetric nuclear matter~\cite{tanihata2013PPNP,ozawa2001nuclear,yamaguchi2011scaling,zhang2024new}.
Fragmentation cross sections to individual nuclides have also been studied, and numerous models have been developed. 
Furthermore, these empirical formulas can parametrize the cross section well and be used for predicting the production rate of nuclides at the border of our knowledge.

Charge pickup reactions, though occurring in the same high-energy heavy-ion interactions, differ fundamentally from typical fragmentation processes, which involve the loss of protons and/or neutrons.
In charge pickup reactions, the projectile gains one or more protons.
At low incident energies, this cross section is dominated by the sequential transfer process~\cite{lenske1989reaction}, which is well described by models like the distorted wave Born approximation (DWBA).
However, the transfer mechanism becomes negligible at high energies ($E/A \gg E_b$, the binding energy), and other distinct processes come into play. 
One is the charge exchange reaction, correlating to the Gamow-Teller (GT) resonance, involving a spin–isospin flip of a nucleon in 
the projectile~\cite{lenske2018theory}.
The other is the excitation of a $\Delta$ resonance and its subsequent decay to a proton and a pion~\cite{gaarde1991isobar,rodriguez2022systematic}. 
This process has also been discussed as a potential solution to the long-standing quenching problem in beta-decay and charge-exchange physics~\cite{lenske2019heavy}.
For consistency with prior work, these reactions are referred to as ``pickup reactions" in this work.

Measurements of charge pickup reaction cross sections ($\sigma_{\mathrm{CP}}$) predominantly focused on stable nuclei at relativistic energies~\cite{olson1981electromagnetic,olson1983factorization,roy1988excitation,bachelier1986first,roy1988excitation,gerbier1988abnormally}. 
Ren, Price, and Williams~\cite{guoxiao1989systematics} proposed an empirical formula for $\sigma_{\mathrm{CP}}$ to describe these existing data: $\sigma_{\mathrm{CP}} = 1.7\times10^{-4}\gamma_{PT}A_P^2$. 
The factor $\gamma_{PT}$, indicating the peripheral nature of collisions, was optimized to $\gamma_{PT}=A_P^{1/3} + A_T^{1/3} - 1.0$, where $A_P$ and $A_T$ are the mass numbers of projectile and target nuclei, respectively.
A power of two dependence on $A_P$ was obtained with surprise. 
With the cascade model combined with statistical evaporation processes, Sümmerer \textit{et al.} suggested that the evaporation stage amplifies the $A_P$ dependence, which is relatively weak in the prefragment stage~\cite{summerer1995charge}. 
This highlights how the neutron number of the projectile affects the final cross sections.

Subsequent measurements of intermediate heavy stable projectiles, such as $^{109}$Ag~\cite{nilsen1994charge} and $^{139}$La~\cite{cummings1990determination},
showed $\sigma_{\mathrm{CP}}$ increase exponentially with the isospin asymmetry, $I_P=(N-Z)/(N+Z)$.
$N$ and $Z$ represent the number of neutrons and protons in the nuclide, respectively.
This trend suggests that neutron-rich nuclides, particularly when studied along isotopic chains, present a promising opportunity to explore the evolution of $\sigma_{\mathrm{CP}}$ with isospin asymmetry.

Despite these insights, data on $\sigma_{\mathrm{CP}}$ for neutron-rich nuclides, particularly at relativistic energies, remain scarce. 
Attempts in this direction include the measurements for $^{18\text{\textendash}20}$C at 40 MeV/nucleon on H target~\cite{yamaguchi2011nuclear}, $^{30,32,33}$Na at 240 MeV/nucleon on H target~\cite{ozawa2014charge}, and $^{12\text{\textendash}19}$C at approximately 900\textit{A} MeV on both C and H targets~\cite{tanihata2016observation}.
In particular, the latter represents a systematic measurement along one isotopic chain and shows a near-exponential growth of $\sigma_{\mathrm{CP}}$ with neutron number.
However, whether this trend is unique to carbon or a general feature of other isotopes remains unclear.

In this work, we present a high-precision measurement of $\sigma_{\mathrm{CP}}$ for several \textit{p}-shell isotopic chains at around 900\textit{A} MeV on both carbon and hydrogen targets.
The study includes 24 isotopes ($^{8,9}$Li, $^{10\text{\textendash}12}$Be, $^{10,13\text{\textendash}15}$B, $^{12,14\text{\textendash}19}$C and $^{14,15,17\text{\textendash}22}$N).
Of them, the data for Li, Be, B, and N isotopes are presented for the first time.
Our findings reveal a rapid increase in $\sigma_{\mathrm{CP}}$ with neutron number, emphasizing a stronger correlation with isospin asymmetry than with mass number. 
A new parametrization explicitly including isospin asymmetry is proposed, providing a unified description for both stable and neutron-rich nuclei.

\section{experiment and result}

\begin{figure}[htbp]
\centering
\includegraphics[width=0.45\textwidth]{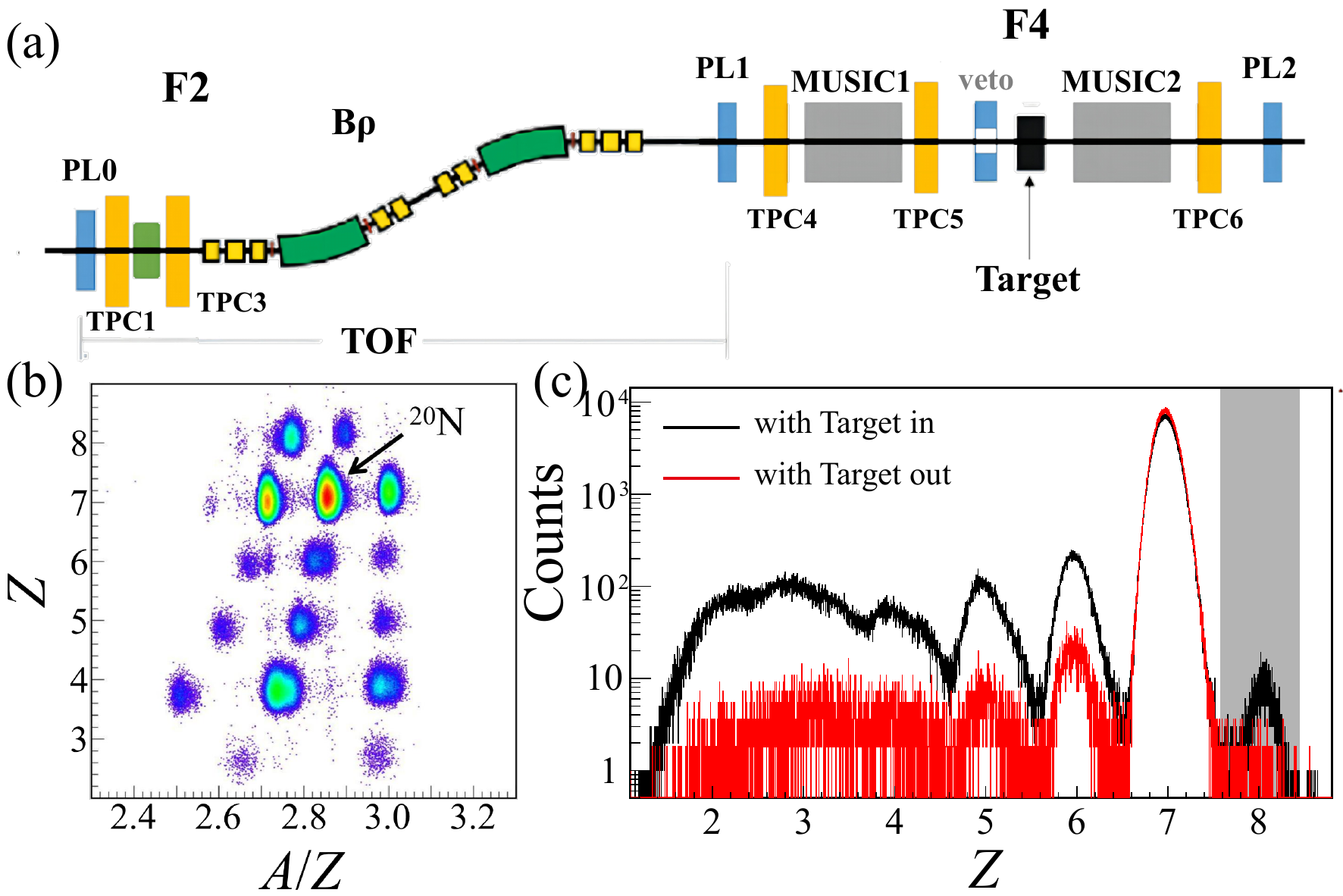}
\caption{\label{fig: PID} 
(a): Schematic view of the experimental setup at the focal plane F2 and F4 in the FRS spectrometer. 
(b): Particle identification spectrum before the reaction target with $^{20}$N indicated by an arrow.
(c) $Z$ identification spectrum after the reaction target with projectile selected as $^{20}$N. 
The black and red histograms are with and without the carbon target, respectively. 
The total incident particle number with the former normalizes the latter.
The shaded area represents the ${Z = 8}$ selection window.
}
\end{figure}

The experiment was conducted using the fragment separator FRS at GSI, Germany~\cite{geissel1992gsi}. 
The isotopes of interest were produced by fragmentations of 1$A$ GeV $^{22}$Ne and $^{40}$Ar ions on a 6.3 g/cm$^2$ Be target, then separated and identified in flight on the event-by-event basis by the magnetic rigidity ($B\rho$), time of flight (TOF), and energy loss (${\Delta}E$) measurements.
TOF measurements were taken using plastic scintillation detectors at the focal plane F2 (PL0) and before the reaction target at F4 (PL1). 
A pair of multi-sampling ion chambers (MUSIC1 and MUSIC2) \cite{stolz2002projectile} were placed before and after the reaction target to measure the energy loss ${\Delta}E$ of the incident and outgoing particles, respectively, providing $Z$ identification. 
The track of an incident particle was determined by a pair of time-projection chambers (TPC4 and TPC5) \cite{janik2011time} located before and after MUSIC1.
This track information was used to select the positions and angles of incident particles on the reaction target.
The position and time information of the reaction production were measured by the TPC6 detector and PL2 detector after MUSIC2, respectively.
The experimental setup, shown in Fig.~\ref{fig: PID}(a), is the same as shown in Ref.~\cite{bagchi2019neutron}.

Charge pickup cross sections $\sigma_{\mathrm{CP}}$ were measured on a 4.01 g/cm$^2$ carbon target and a 3.375 g/cm$^2$ polyethylene target. 
Fig.~\ref{fig: PID}(b) shows a typical particle identification example using $^{20}$N projectile.
The number of incident nuclei ($N_\mathrm{in}$) was determined based on TOF, ${\Delta}E$ by MUSIC1, and the incident position and angle.
The resolution of MUSIC is approximately $0.12Z$ ($\sigma$) and the estimated contamination level of other nuclei in the projectile is less than $0.1\%$.
The atomic number \textit{Z} of nuclei after the target was determined from the pulse height spectrum of the MUSIC2. 
Fig.~\ref{fig: PID}(c) shows a typical \textit{Z}-identification spectrum for $^{20}$N incident,
with the black and red histograms representing data with and without the carbon target, respectively.
To obtain the number of nuclei of the charge pickup ($N_{\Delta Z = + 1}$), a selection window was placed around the $Z = 8$ peak on both sides of the spectrum (shaded area in Fig.~\ref{fig: PID}(c)).
To avoid contamination from ${Z = 7}$, the lower limit for the selection window was set at ${Z = 7}$ peak centroid plus 5$\sigma$, where $\sigma$ was obtained via Gaussian fitting.
The upper limit was set at $Z = 8$ peak centroid plus 4$\sigma$, where $\sigma$ is the average of $Z =$ 5, 6, and 7 peaks.
The peak position of $Z=8$ was determined by the target-in spectrum and also applied to the target-out spectra.
Taking $^{20}$N as an example, the contamination from ${Z = 7}$ events is estimated to be only $2.2 \times 10^{-5}$ of the total charge pickup events.
The fractions of true ${Z = 8}$ events excluded by the lower and upper bounds are estimated to be less than $2.3 \times 10^{-4}$ and $3.2 \times 10^{-5}$. Therefore, the selection window is safe to guarantee the precise counting of $Z=8$ events. 
Furthermore, the detectors after the target cover a large enough solid angle, ensuring that most of the ${\Delta Z = + 1}$ fragments are detected, making the missing cross section negligibly small.

The charge pickup cross sections are obtained by,
\begin{equation} \label{eq:calcs}
  \sigma_{\mathrm{CP}} = t^{-1}(R_\mathrm{T}-R_\mathrm{0}), 
\end{equation}
where $R=N_{\Delta Z=+1}/ N_\mathrm{in}$, and $R_\mathrm{T}$ and $R_\mathrm{0}$ refer to measurements with and without the reaction target, respectively.
The number of the target nuclei in the unit of cm$^{-2}$ is shown as \textit{t}.
The cross sections on the proton were derived by subtracting the C target cross sections from those of the polyethylene target.
A reanalysis of $\sigma_{\mathrm{CP}}$ for carbon isotopes from Ref.~\cite{tanihata2016observation} is performed by considering the impact of the veto detector before the target, as detailed in Ref.~\cite{zhang2024new}.
The obtained charge pickup cross sections on the carbon and hydrogen targets are presented in Table~\ref{tab:Cross}.

\begin{table}[htbp]
\centering
\caption{\label{tab:Cross} Charge pickup reaction cross section of 24 \textit{p}-shell isotopes on carbon
and hydrogen targets.
}
\begin{tabular}{c c c c }
\hline
\multirow{2}*{Isotope} & \textit{E}/\textit{A} & C target & H target \\
~ & MeV & mb & mb \\
\hline
$^{	8	}$Li &	901 	&	$1.7 \pm 1.0$&	$0.4 \pm 1.4$\\
$^{	9	}$Li &	958 	&	$2.0 \pm 0.5$&	$1.5 \pm 0.5$\\
$^{	10	}$Be &	994 	&	$2.0 \pm 0.8$&	$0.5 \pm 0.7$\\
$^{	11	}$Be &	928 	&	$2.4 \pm 0.7$&	$0.4 \pm 0.6$\\
$^{	12	}$Be &	959 	&	$2.9 \pm 0.4$&	$1.8 \pm 0.4$\\
$^{	10	}$B	&	930 	&	$0.0 \pm 0.5$&	$0.02 \pm 0.4$\\
$^{	13	}$B	&	933 	&	$0.8 \pm 0.5$&	$0.5 \pm 0.6$\\
$^{	14	}$B	&	990 	&	$1.7 \pm 0.8$&	$0.6 \pm 0.8$\\
$^{	15	}$B	&	963 	&	$3.2 \pm 0.2$&	$1.8 \pm 0.2$\\

$^{	12	}$C	&	928 	&	$0.0 \pm 0.1$&	$0.1 \pm 0.1$\\
$^{	14	}$C	&	991 	&	$0.4 \pm 0.3$&	$0.6 \pm 0.3$\\
$^{	15	}$C	&	893 	&	$0.9 \pm 0.1$&	$0.7 \pm 0.1$\\
$^{	16	}$C	&	808 	&	$2.8 \pm 1.1$&	$2.2 \pm 1.1$\\
$^{	17	}$C	&	962 	&	$3.5 \pm 0.4$&	$1.9 \pm 0.4$\\
$^{	18	}$C	&	955 	&	$6.1 \pm 0.3$&	$3.7 \pm 0.3$\\
$^{	19	}$C	&	880 	&	$8.2 \pm 0.6$&	$3.8 \pm 0.5$\\

$^{	14	}$N	&	924 	&	$0.0 \pm 0.1$&	$0.02 \pm 0.1$\\
$^{	15	}$N	&	762 	&	$1.5 \pm 1.2$&	$0.5 \pm 1.4$\\
$^{	17	}$N	&	927 	&	$1.0 \pm 0.2$&	$0.3 \pm 0.2$\\
$^{	18	}$N	&	848 	&	$1.6 \pm 0.6$&	$1.2 \pm 0.6$\\
$^{	19	}$N	&	949 	&	$3.3 \pm 0.1$&	$2.1 \pm 0.2$\\
$^{	20	}$N	&	877 	&	$5.1 \pm 0.6$&	$2.9 \pm 0.7$\\
$^{	21	}$N	&	874 	&	$7.3 \pm 0.9$&	$4.7 \pm 0.8$\\
$^{	22	}$N	&	882 	&	$8.5 \pm 1.0$&	$6.7 \pm 1.0$\\
\hline
\end{tabular}
\end{table}

\section{discussion}

First, let us compare $\sigma_{\mathrm{CP}}$ between stable and neutron-rich unstable nuclei.
We surveyed the existing data with reaction energies from 0.7 to 2.1 GeV/nucleon. 
Sixteen reactions~\cite{olson1981electromagnetic,olson1983factorization,gerbier1988abnormally,guoxiao1989systematics,cummings1990determination,nilsen1994charge,summerer1995charge} with high precision and unambiguous descriptions are selected for the following analyses,
covering stable projectile nuclei with mass numbers between 12 and 197 on various targets.
We focus on such energy domains because, at relativistic energies, charge pickup reactions predominantly proceed via a charge-exchange process, minimizing contributions from low-energy mechanisms such as sequential transfer, and, thereby, exhibiting a minor dependence on the incident energy.
Additionally, the experimental data below 700 MeV/nucleon are sparse and display markedly higher cross sections, complicating a consistent comparison. 
At above 2.1 GeV/nucleon, systematic measurements are currently unavailable.

\begin{figure}[htb]
\centering
\includegraphics[width=0.45\textwidth]{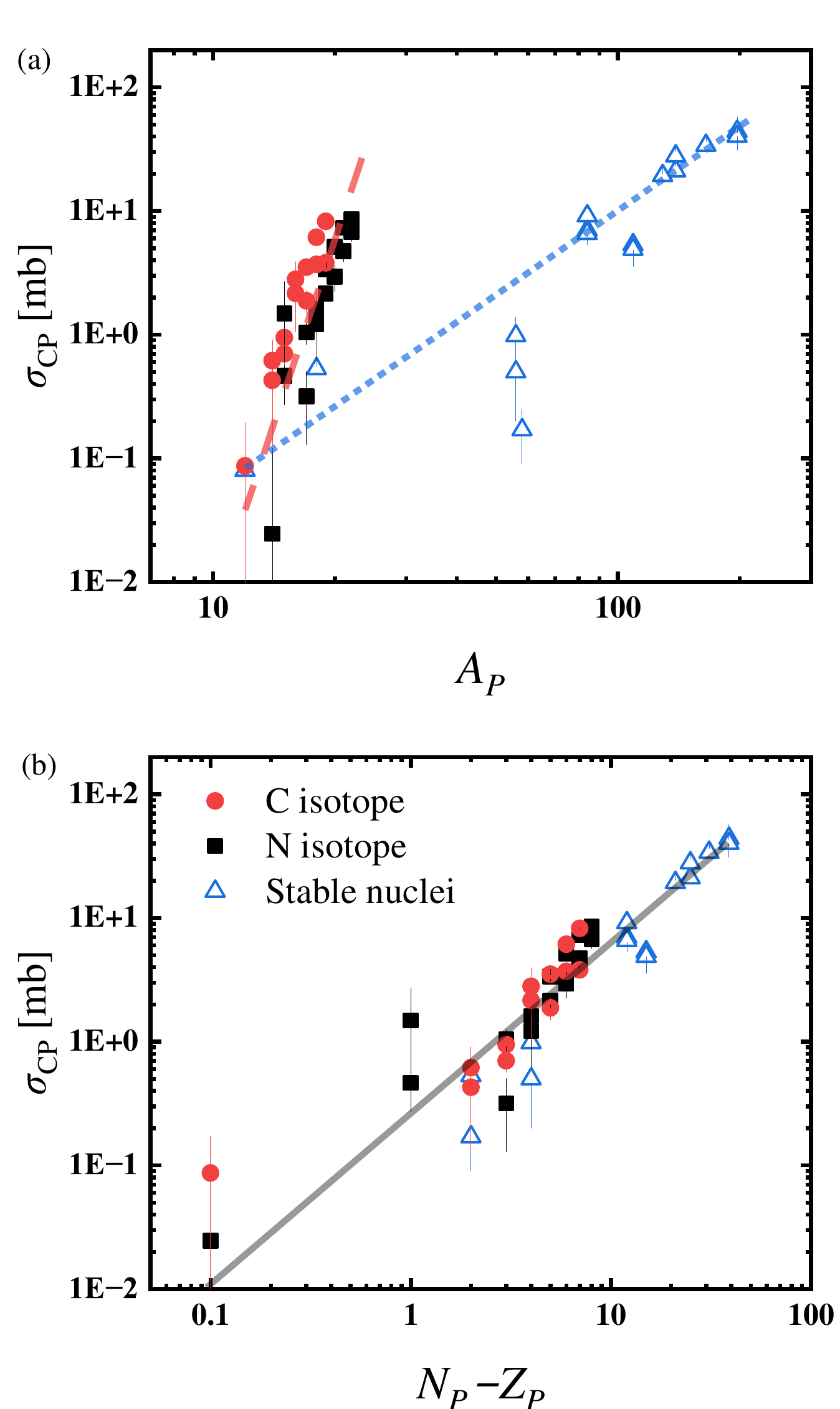}
\caption{\label{fig:compare} 
Charge pickup cross sections for both $Z \geq 6$ stable nuclei and unstable (neutron-rich) nuclei at relativistic energy on several targets are plotted against the mass number (a), and the neutron excess $N_P-Z_P$ (b). 
The blue dotted line in panel (a) represents the empirical formulas $\sigma_{\mathrm{CP}}=1.7\times10^{-4}\gamma_{PT}A_P^2$.
Nuclides with $N_P-Z_P = 0$ in panel (b) are shifted to 0.1 for better visibility. 
The red dashed line in (a) and the black solid line in (b) are drawn to guide the eye. 
}
\end{figure}

The $\sigma_{\mathrm{CP}}$ of compiled data and our reported $Z \ge 6$ unstable nuclei are plotted together in Fig.~\ref{fig:compare}(a).
The trends in the cross sections involving stable and unstable nuclides for mass number $A_P$ are depicted by the blue dotted and the red dashed lines, respectively. 
Despite the limited overlap between the two datasets, the varying slopes still suggest that the cross sections are not a mere function of the mass number of the projectile.
Note that $^{56}$Fe~\cite{guoxiao1989systematics,cummings1990determination} and $^{58}$Ni~\cite{cummings1990determination} deviate significantly from the empirical formula.
In contrast, when the same cross sections are plotted against neutron excess of projectile nuclei, $N_P - Z_P$, we observe a consistent trend across all the stable and unstable nuclei, as in Fig.~\ref{fig:compare}(b). 
This indicates that the neutron excess is a more appropriate parameter for discussing the charge pickup reaction cross sections of atomic nuclei globally.
It is worth noting that the existing empirical formula cannot reproduce the new data involving the light unstable isotopes. 
This highlights the need for a consistent description of stable and unstable isotopes.

For simplicity, we first consider the reactions with a proton target. 
In our measurement, we observe a proton pickup from the projectile without determining its mass.
Consequently, (\textit{p}, \textit{n}) exchange reactions occurring below the proton emission threshold—typically the dominant process at relativistic energies~\cite{lenske1989reaction}—are fully integrated into the measured $\sigma_{\mathrm{CP}}$, regardless of the following neutron emissions.
It includes all the transitions governed by Fermi and Gamow-Teller selection rules, which dominate at small scattering angles and high energies~\cite{fujita2011spin}, and transitions governed by other high-order selection rules.

\begin{figure}[htbp]
\centering
\includegraphics[width=0.45\textwidth]{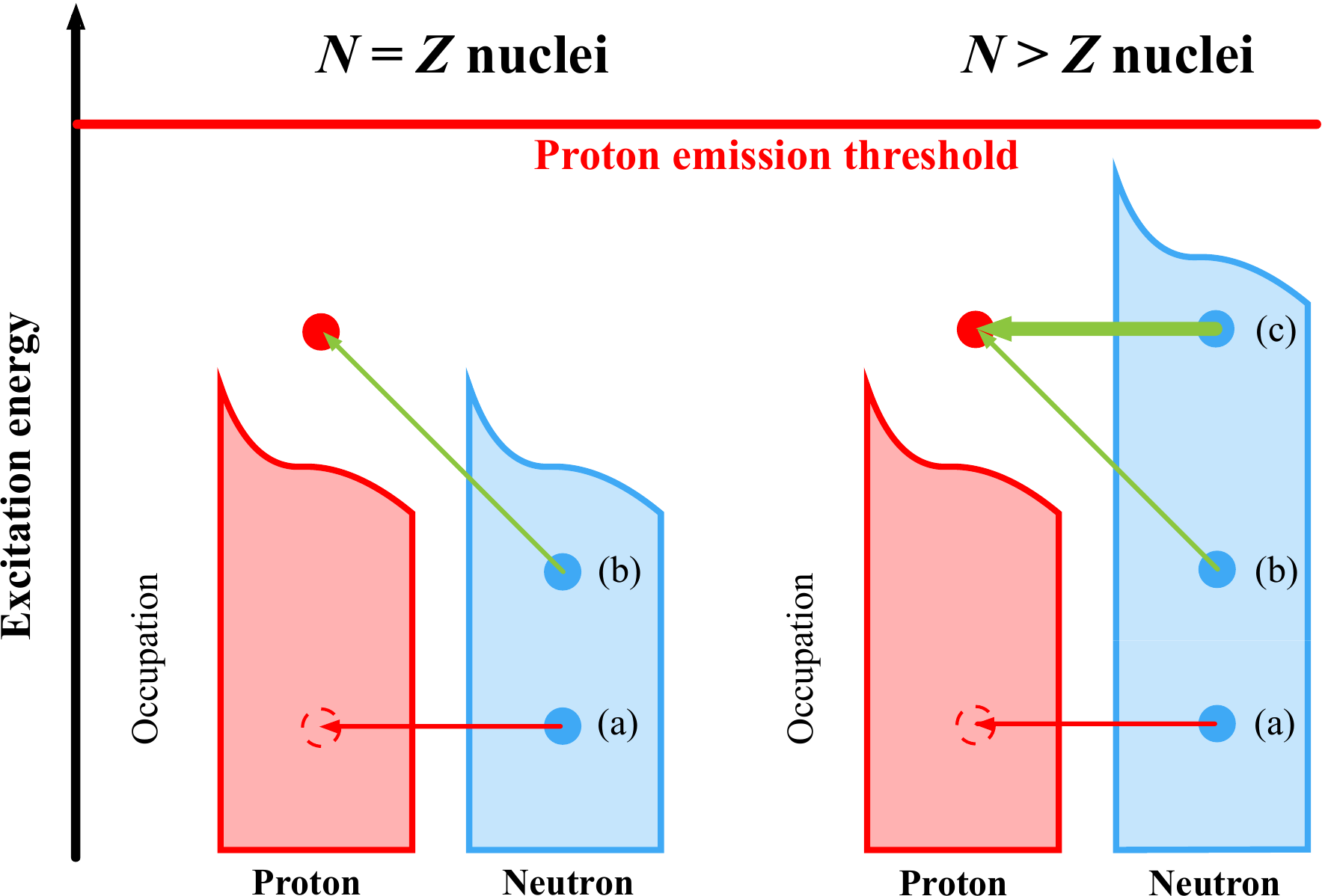}
\caption{\label{fig:occupation} The occupation and neutron to proton transition configuration of $N=Z$ and $N>Z$ nuclei. For details, refer to the text. 
}
\end{figure}

In this context, it is natural to consider the cross sections strongly related to the neutron excess, as illustrated in Fig.~\ref{fig:occupation}.
The figure depicts the occupation configurations of protons and neutrons in single-particle states for both $N=Z$ and $N>Z$ nuclei.
A (\textit{p}, \textit{n}) transition, expressed by an arrow, indicates the exchange from a neutron to a proton.
For $N=Z$ nuclei, protons and neutrons occupy the same orbitals, preventing the superallowed transitions (transitions within the same orbitals), as shown in type (a) of the figure.
Transitions can occur only for neutrons near the occupation limit, as type (b) indicates, but these are relatively small because superallowed transitions are not included.
In contrast, for $N>Z$ nuclei, the superallowed transitions are abundantly possible because there are more proton orbitals available, as shown in type (c).
The number of available orbitals is directly related to the neutron-proton difference ($N_P-Z_P$).

We parametrize the cross sections explicitly, including neutron excess. 
First of all, we factorize the cross section as:
\begin{equation} \label{eq: factorized}
\sigma_{\mathrm{CP}}(P,T) = a \: \gamma_{PT} \: F(P) \quad \mathrm{[mb]}.
\end{equation}
Here, $\gamma_{PT}$ presents the peripheral nature of collisions between the target and projectile nuclei:
\begin{equation} \label{eq: gammaPT}
 \gamma_{PT} = A_P^{1/3} + A_T^{1/3} - c.
\end{equation}
This parametrization is chosen for historical reasons.

The function $F(P)$, which characterizes the projectile nucleus, should be factorized into two components.
The first component, as given in Ref.~\cite{guoxiao1989systematics}, is the mass number dependence expressed as $A_P^m$.
The second component is required to characterize the effect induced by the isospin asymmetry $I_P$, defined as 
$({N_P - Z_P})/({N_P + Z_P})$.
We present the form for $F(P)$ as:
\begin{equation} \label{eq: FNP}
 F(P) = A_P^m \; \left({1 + b\; I_P}\right) ^n,
\end{equation}
where \textit{b} is a tuning parameter.
The coefficients \textit{a}, \textit{b}, and \textit{c}, along with the power numbers \textit{m} and \textit{n}, were determined via the least-squares method. The best-fit values are presented in Table~\ref{tab:Constants}.
Hereafter, we describe the procedure for determining the parameters.

\begin{table}[htbp]
\centering
\caption{\label{tab:Constants} Constants used in the new charge pickup cross-section formula}
\begin{tabular}{l c c }
\hline
Parameter & Constant & Value \\
\hline
Scaling factor & \textit{a} & 0.0014 $\pm$ 0.0003 \\
Isospin asymmetry tuning & \textit{b} & 2.5 $\pm$ 1.4 \\
Overlap parameter in collision & \textit{c} & 1.42 $\pm$ 0.17 \\
Mass number dependence & \textit{m} & 1.12 $\pm$ 0.03 \\
Isospin asymmetry dependence & \textit{n} & 6 $\pm$ 2 \\

\hline
\end{tabular}
\end{table}

\subsection{Target dependence}

\begin{figure}[htbp]
\centering
\includegraphics[width=0.4\textwidth]{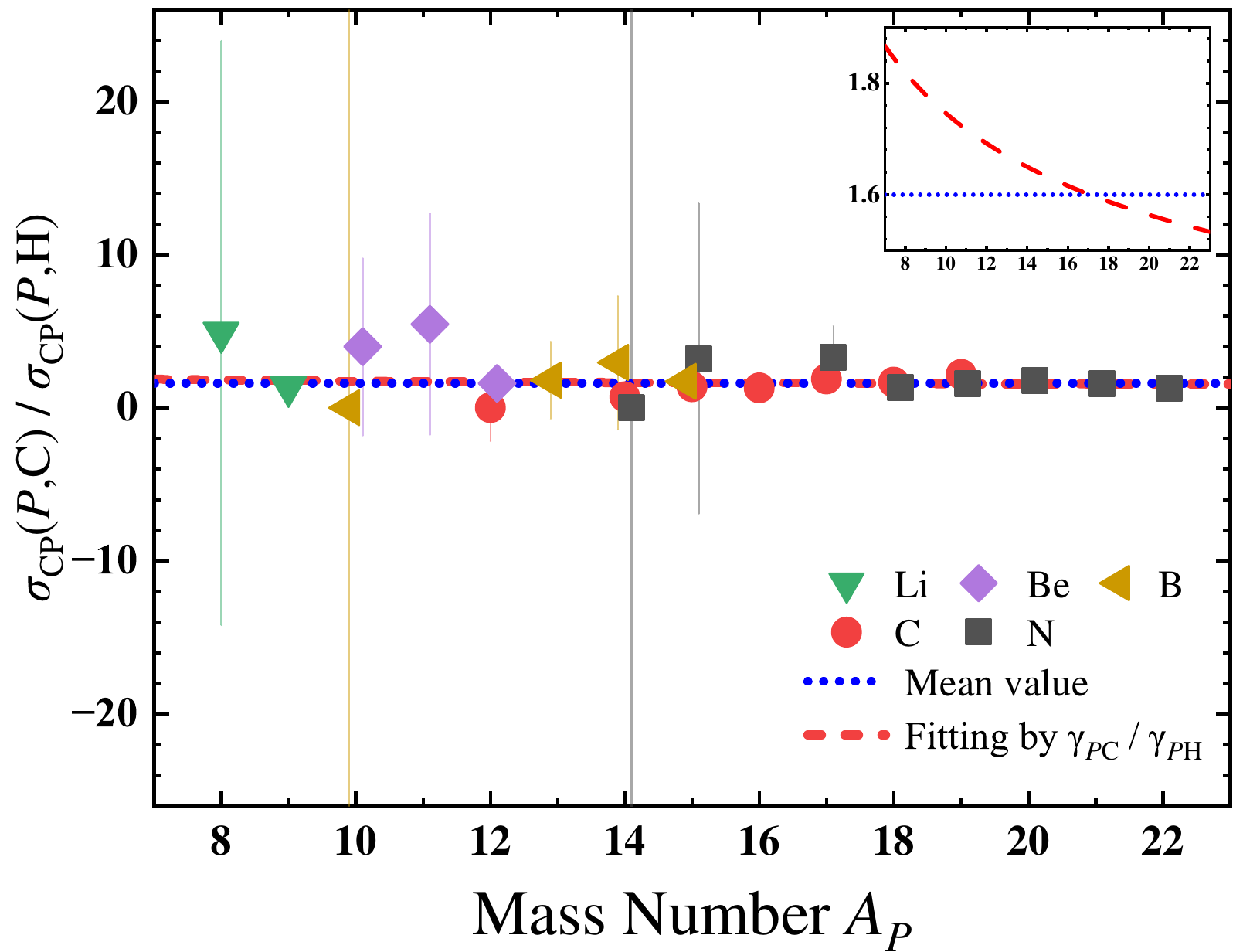}
\caption{\label{fig:Sigmaratio} 
Ratios of $\sigma_{\mathrm{CP}}(P,\mathrm{C})$ to $\sigma_{\mathrm{CP}}(P,\mathrm{H})$ as a function of projectile’s mass number $A_P$. 
The blue dotted line and the red dashed line represent the mean value and best fit of the cross-section ratios, respectively. 
The inset shows the same plot but focusing on the two lines.
The values of the mass number for the data have been slightly shifted for better visibility.
}
\end{figure}

To quantify the target dependence, we compared the cross sections of the C and H targets in detail.
The radios of the cross sections are
\begin{equation} \label{eq: sigma_ratio}
  \frac{\sigma_{\mathrm{CP}}(P,\mathrm{C})}{\sigma_{\mathrm{CP}}(P,\mathrm{H})} \;,
\end{equation}
where $\sigma_{\mathrm{CP}}(P,T)$ indicate the charge pickup cross sections of projectile \textit{P} and target \textit{T}. 
The ratios are almost constant as shown in Fig.~\ref{fig:Sigmaratio}.
It suggests that the difference in $\sigma_{\mathrm{CP}}$ between different targets for all nuclides, including unstable nuclei, can be well factorized with Eq.~\ref{eq: factorized}.

We then kept the target factor in $\gamma_{PT}$ defined by Eq.~\ref{eq: gammaPT}. 
Then the ratio is written as,
\begin{equation} \label{eq: gamma_ratio}
  \frac{\sigma_{\mathrm{CP}}(P,\mathrm{C})}{\sigma_{\mathrm{CP}}(P,\mathrm{H})} = \frac{\gamma_{P\mathrm{C}}}{\gamma_{P\mathrm{H}}} = \frac{A_P^{1/3} + 12^{1/3} - c}{A_P^{1/3} + 1^{1/3} - c} \;.
\end{equation}
The overlap parameter \textit{c} is considered to be constant for all reactions due to the minor energy dependence of $\sigma_{\mathrm{CP}}$ in the relativistic energy region~\cite{kelic2004isotopic}. 
Applying the least-squares method to fit Eq.~\ref{eq: gamma_ratio}, \textit{c} is determined to be 1.42(17).
Although the constant $c$ is not strictly equivalent to a constant ratio of the cross sections, the resulting difference is negligible, 
as illustrated by the inset in Fig.~\ref{fig:Sigmaratio}.
Our overlap value is more significant than Ref.~\cite{guoxiao1989systematics} ($c = 1.0$). 
This discrepancy is attributed to two datasets characterized by distinct energies, projectiles, and targets. 
However, it has minimal impact on the subsequent factorization of the charge pickup cross sections. Moreover, 
Eq.~\ref{eq: gamma_ratio}, which describes the cross-section ratios well, shows a weak dependence on the size of the target nucleus and points to the collision being very peripheral.

\subsection{Projectile dependence}

As mentioned earlier, the function $F(P)$ representing the nature of the projectile includes two orthogonal terms: the mass number $A_P$ and the isospin asymmetry $I_P$.
We consider using the exponential and power functions of $I_P$.
The exponential type function is: $\sigma_{\mathrm{CP}}=a\gamma_{PT}A_P^m e^{nI_P}$, similar to Ref.~\cite{nilsen1994charge}.
A simple power-type function $\sigma_{\mathrm{CP}}=a\gamma_{PT}A_P^mI_P^n$ poses a problem for $N_P \leq Z_P$ due to its negative value.
To rectify this, we select a modified power function $\sigma_{\mathrm{CP}}=a\gamma_{PT}A_P^{m}(1+bI_P)^n$ by introducing a tuning parameter $b$, which determined through fits.

Our new results and the existing data, excluding Fe and Ni, 
were fitted to either the exponential function or the modified power function via the least-squares method. 
The reduced chi-square ($\chi^2_v$) and the values of each parameter $m$, $n$, $a$, and $b$ determined by the fit for both functional types are presented in Table~\ref{tab:Parameter}. 
Including Fe and Ni data would result in a slightly increased $\chi^2_v$ to 6.0 for the exponential function and 3.4 for the modified power function.

\begin{table*}[htbp]
\centering
\caption{\label{tab:Parameter} Fitting results and regression model accuracy metrics for different function types.}
\begin{tabular}{l c c c c c }
\hline 
From & Reduced chi-square & \textit{m} & \textit{n} & \textit{a} & \textit{b} \\
\hline
$a\gamma_{PT}A_P^m e^{nI_P}$ & 2.8 & 1.18 $\pm$ 0.03 & 10.1 $\pm$ 0.3 & 0.0017 $\pm$ 0.0003 & --- \\
$a\gamma_{PT}A_P^{m}(1+bI_P)^n$ & 2.5 & 1.12 $\pm$ 0.03 & 6 $\pm$ 2 & 0.0014 $\pm$ 0.0003 & 2.5 $\pm$ 1.4 \\
\hline
\end{tabular}
\end{table*}

\begin{figure*}[htbp]
\centering
\includegraphics[width=0.8\textwidth]{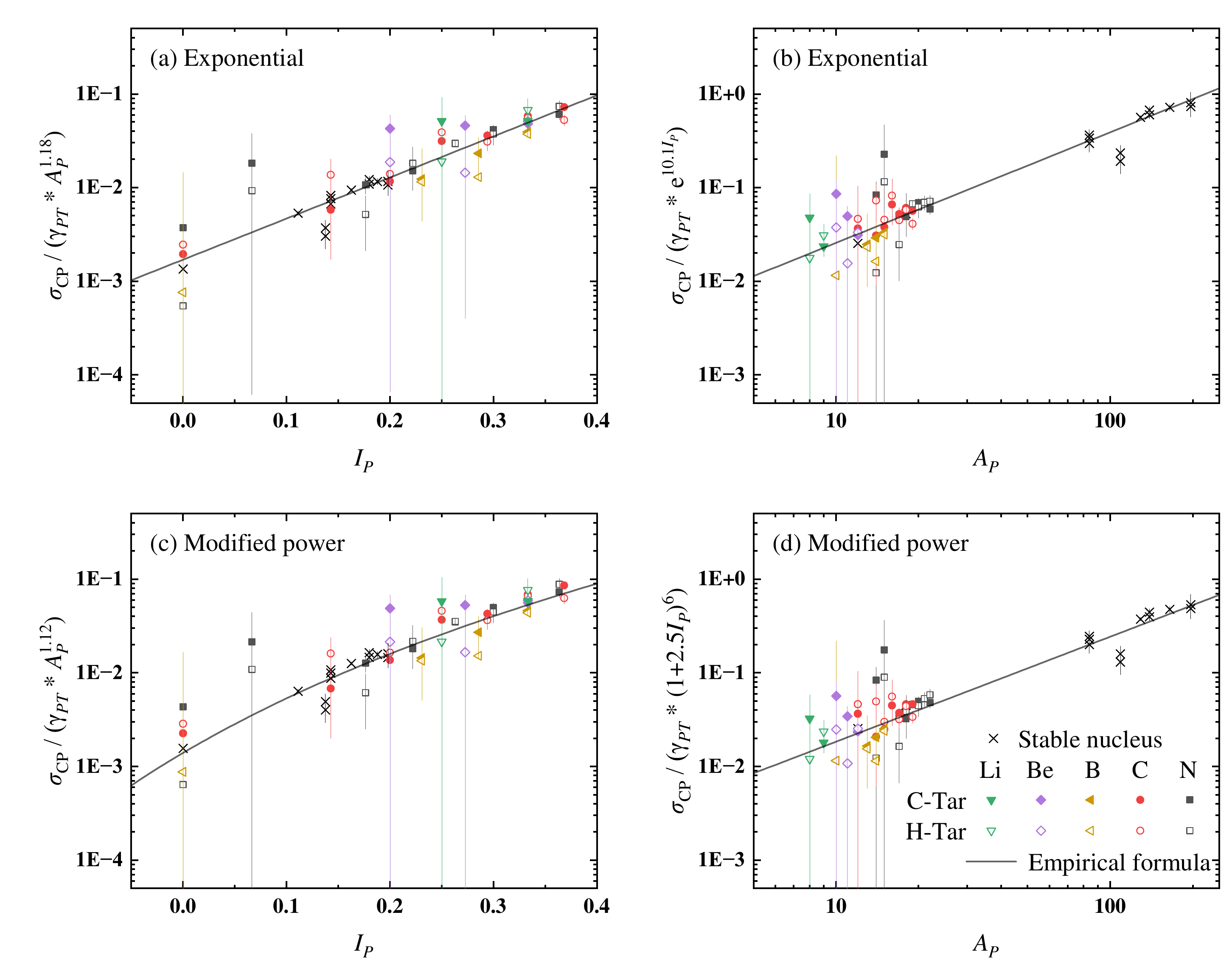}
\caption{\label{fig:twoterms}
The normalized experiment cross section $\sigma_{\mathrm{CP}}/(\gamma_{PT}A_P^{1.18})$ against $I_P$ (a), $\sigma_{\mathrm{CP}}/(\gamma_{PT}\textit{e}^{10.1I_P})$ against $A_P$ (b) obtained from the exponential function. And $\sigma_{\mathrm{CP}}/(\gamma_{PT}A_P^{1.12})$ against $I_P$ (c) and $\sigma_{\mathrm{CP}}/(\gamma_{PT}(1+2.5I_P)^6)$ against $A_P$ (d) obtained from the modified power function.
The representation of colored symbols is consistent with Fig.~\ref{fig:Sigmaratio}.
The black cross symbols represent the existing stable nuclei in the literature~\cite {olson1981electromagnetic,olson1983factorization,guoxiao1989systematics,gerbier1988abnormally,cummings1990determination,nilsen1994charge,summerer1995charge}. 
The solid black lines in (a) and (b) represent the empirical exponential function corresponding to the first row in Table~\ref{tab:Parameter}, and in (c) and (d) represent the modified power function corresponding to the second row in Table~\ref{tab:Parameter} (i.e., Eq.~\ref{eq: final}).
}
\end{figure*}

In Fig.~\ref{fig:twoterms}, we compare the two fitting formulations presented in Table~\ref{tab:Parameter}.
The normalized experiment cross section $\sigma_{\mathrm{CP}}/(\gamma_{PT}A_P^{1.18})$ against $I_P$ and $\sigma_{\mathrm{CP}}/(\gamma_{PT}\textit{e}^{10.1I_P})$ against $A_P$ obtained from the exponential function are shown in Fig.~\ref{fig:twoterms}(a) and (b), respectively.
The $\sigma_{\mathrm{CP}}/(\gamma_{PT}A_P^{1.12})$ against $I_P$ and $\sigma_{\mathrm{CP}}/(\gamma_{PT}(1+2.5I_P)^6)$ against $A_P$ obtained from the modified power function are shown in Fig.~\ref{fig:twoterms}(c) and (d), respectively. 
Both equations yield similar predictions for charge pickup cross sections and perform well for both $I_P$ and $A_P$ dependence.
As shown in Fig.~\ref{fig:twoterms}(a) and (c), the predictions align well with the observations for both neutron-rich nuclides ($I_P>0.2$) and heavy stable nuclides ($0.2>I_P>0.1$). 
The values are also consistent within errors for light stable nuclei ($I_P<0.1$), suggesting that the predictions accurately represent the dependence of the isospin asymmetry.
Additionally, as shown in Fig.~\ref{fig:twoterms}(b) and (d), the empirical formula effectively describes the dependence on $A_P$ and fits well with the data on both sides of the mass number.

\begin{figure*}[htbp]
\centering
\includegraphics[width=0.8\textwidth]{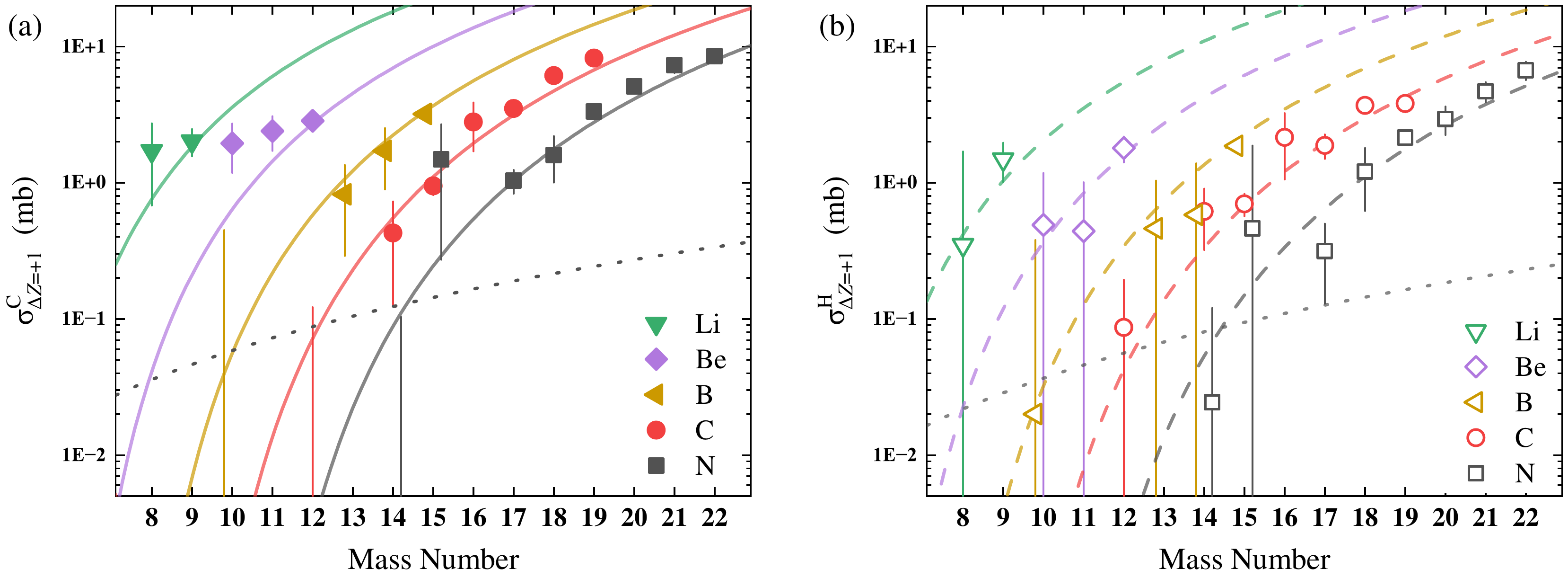}
\caption{\label{fig:Sigmaallratio} Charge pickup cross sections of 24 isotopes on the C (a) and H target (b). 
The solid and dashed lines represent the predictions using our new empirical formulas for the cross sections on the C and H targets, respectively, with the previous empirical formulas $\sigma_{\mathrm{CP}}=1.7\times10^{-4}\gamma_{PT}A_P^2$ plotted as dotted lines for comparison.
}
\end{figure*}

\subsection{Empirical formula}

We choose the modified power function of reduced chi-square that is closest to 1, as our new empirical formula:
The fit metric of the exponential type function is comparable to, but marginally poorer than, that of the modified power type.
We have also assessed other forms expressing neutron excesses, such as mathematical variants like $N_P/A_P$ and $N_P/Z_P$, or quantities associated with the empirically smoothed stable line.
However, the accuracy of these forms is worse than that of $I_P$ according to the fit metrics.
The expression of the proposed formula is:
\begin{equation} \label{eq: final}
\sigma_{\mathrm{CP}} = 0.0014\gamma_{PT} A_P^{1.12}\left({1 + 2.5\frac{ N_P - Z_P}{N_P + Z_P}}\right) ^{6} \mathrm{[mb]}.
\end{equation}
Compared to the $A_P^2$ formula, our new formula exhibits a weaker dependence on the mass number, reducing the exponent from 2 to 1.12. 
For stable nuclides, the new formula still predicts well because the isospin asymmetry $I_P$ of heavy nuclei contributes more significantly than those of light nuclei, partially compensating for the reduced $A_P$ dependence. 
The new formula is highly sensitive to the isospin asymmetry, with a power of 6.
For instance, from $^{12}$C to $^{19}$C, the value of $I_P$ increases from 0 to 0.37, resulting in a 50-fold increase in the corresponding cross section.
This high sensitivity can be explained by the increase in the number of neutrons in the projectile, especially when the last neutron orbit reaches the \textit{sd} shell, significantly increasing the number of possible transitions from one neutron to one proton during the reaction~\cite{tanihata2016observation}, as shown in Fig.~\ref{fig:occupation}.

Fig.~\ref{fig:Sigmaallratio} displays the measured cross sections alongside predicted values given by Eq.~\ref{eq: final}.
The empirical formula accurately replicates the experimental data $\sigma_{\mathrm{CP}}$ along each isotopic chain and applies to all reaction targets.
A notable discrepancy is observed in the cross section of $^{10,11}$Be on the C target.
Although the exact cause remains unknown, the predicted values are within twice the error range of the experimental values, making this acceptable.
The combined predictions for stable and neutron-rich nuclei demonstrate that our new formula successfully decouples and describes the dependence of the charge pickup cross section on both the projectile’s mass number and isospin symmetry for the first time.
This advancement will facilitate a fast and reliable evaluation of production rates in most exotic neutron-rich nuclides.

The empirical formula presented here was derived and validated within the relativistic energy region of 0.7-2.1 GeV/nucleon, where the reaction mechanism is primarily dominated by charge-exchange processes and the energy dependence of the cross sections is minor.
While we expect the formula to remain applicable at several GeV/nucleon, the absence of experimental data above 2.1 GeV/nucleon currently prevents its direct verification.
Extending this parametrization to lower energies is interesting.
However, the data below 700 MeV/nucleon are scarce and limited to a few heavy stable nuclei around 500 MeV/nucleon, as reported in Refs.~\cite{nilsen1994charge,cummings1990determination}.
These data show a significant increase in cross sections, indicating the emergence of different reaction mechanisms compared to those above 700 MeV/nucleon. 
Therefore, the lower energy limit of the current parametrization is set to 700 MeV/nucleon.
Future accurate measurements at a wider energy domain are essential to validate further and possibly extend our parametrization.

\section{summary}

We measured the charge pickup reaction cross sections for 24 \textit{p}-shell isotopes, including $^{8,9}$Li, $^{10\text{\textendash}12}$Be, $^{10,13\text{\textendash}15}$B, $^{12,14\text{\textendash}19}$C and $^{14,15,17\text{\textendash}22}$N, at around 900$A$ MeV on both carbon and hydrogen targets.
Notably, the measurements for Li, Be, B, and N isotopes are reported here for the first time.
By comparing data from the two targets, we redetermined the collision parameter $\gamma_{PT}= A_P^{1/3}+A_T^{1/3}-1.42$ for the charge pickup cross section. We proposed an empirical formula as $\sigma_{\mathrm{CP}} = 0.0014\gamma_{PT} A_P^{1.12} (1+2.5I_P)^{6} \mathrm{[mb]}$, where $I_P=(N_P-Z_P)/(N_P+Z_P)$.
The power of two in mass dependence in the previous studies is significantly reduced when using our new systematic data with large isospin asymmetries. 
The new empirical formula, separating the dependence on mass and isospin, highlights the critical role of the latter. It provides a global but precise estimation of the charge pickup cross section.
This will facilitate the evaluation of production rates in most exotic neutron-rich nuclides.

The exponential function has achieved a similar $\chi^2_v$ value as the modified power function. This means that the exact form of the empirical formula may change when more measurements, especially in the neutron-deficient, \textit{sd}- and \textit{pf}-shell nuclides, are supplemented to the existing dataset.
Regardless of the form, the pronounced dependence on the extracted isospin asymmetry has revealed fundamental physical mechanisms underlying the charge pickup process.
As we discussed, during the extremely peripheral collision, the number of neutrons in the projectile exceeding the number of protons exponentially pushes up the quantity of potential neutron-to-proton transitions,
which directly impacts the probability of the charge pickup reaction.
Developing a method for identifying and using this quantity, such as empirical energy-level densities, may indicate a possible direction for improvement in the current empirical formula.
Correspondingly, another perspective is that correlating charge pickup cross sections with energy-level densities might provide a pathway to constrain total transition strengths, either through direct measurement or prediction of cross sections.

\section{acknowledgments}
\begin{acknowledgments}
The authors are thankful for the support of the GSI accelerator staff and the FRS technical staff for efficiently preparing the experiment setup.
The current analysis was performed in FAIR Phase-0. 
This work was partly supported by the National Natural Science Foundation of China (Nos. 12325506, 11961141004) and the 111 Center (No. B20065).
The support from NSERC, Canada, for this work is gratefully acknowledged. 
\end{acknowledgments}

\end{document}